\documentclass[12pt]{iopart}

\usepackage[plainpages=false,breaklinks=true]{hyperref}
\usepackage{graphicx,color}
\usepackage{pdfsync}
\usepackage{tabularx}
\usepackage{ amssymb}
\usepackage{iopams}
\usepackage{color}

\newcommand{\calv}{C_{\mathrm{A}}}
\newcommand{\calva}{\bar C_{\mathrm{A}}}
\newcommand{\cart}{C_{\mathrm{a}}}
\newcommand{\cven}{C_{\bar{\mathrm{v}}}}
\newcommand{\cinh}{C_{\mathrm{I}}}
\newcommand{\cmeas}{C_{\mathrm{meas}}}
\newcommand{\cper}{C_{\mathrm{gut}}}
\newcommand{\crpt}{C_{\mathrm{rpt}}}

\newcommand{\lrpt}{\lambda_{\mathrm{b:rpt}}}
\newcommand{\lper}{\lambda_{\mathrm{b:gut}}}
\newcommand{\hen}{\lambda_{\mathrm{b:air}}}

\newcommand{\qper}{q_{\mathrm{gut}}}
\newcommand{\qalv}{\dot{V}_{\mathrm{A}}}
\newcommand{\qalva}{\bar{ \dot{V}}_{\mathrm{A}}}
\newcommand{\qc}{\dot{Q}_{\mathrm{c}}}

\newcommand{\prl}{k_\mathrm{pr}^{\mathrm{rpt}}}
\newcommand{\prm}{k_\mathrm{pr}^{\mathrm{gut}}}
\newcommand{\ml}{k_\mathrm{met}^{\mathrm{rpt}}}

\newcommand{\valv}{\tilde{V}_{\mathrm{A}}}
\newcommand{\vrpt}{\tilde{V}_{\mathrm{rpt}}}
\newcommand{\vper}{\tilde{V}_{\mathrm{gut}}}

\newcommand{\tb}{\mathbf{p}}

\newcommand{\cb}{\mathbf{c}}

\newcommand{\gb}{\mathbf{g}}

\newcommand{\nulb}{\mathbf{0}}

\usepackage{float}

 \newcommand{\cperb}{C_{\mathrm{gut,b}}}
\newcommand{\crptb}{C_{\mathrm{rpt,b}}}

\newcommand{\vc}{\mathbf{c}}
\newcommand{\vb}{\mathbf{b}}
\newcommand{\vg}{\mathbf{g}}

\makeatletter
\def\dynscriptsize{\check@mathfonts\fontsize{\sf@size}{\z@}\selectfont}
\makeatother
\def\textunderset#1#2{\leavevmode
  \vtop{\offinterlineskip\halign{
    \hfil##\hfil\cr\strut#2\cr\noalign{\kern-.3ex}
    \hidewidth\dynscriptsize\strut#1\hidewidth\cr}}}

\begin{document}

\title[Modeling of breath methane concentration profiles]{Modeling of breath methane concentration profiles
during exercise on an ergometer}

\author[Szab\' o et al.\ ]{Anna Szab\' o$^{1,2}$,  Karl Unterkofler$^{2,3}$,  Pawel Mochalski$^2$, Martin Jandacka$^{2,3}$, Vera Ruzsanyi$^{2,7}$, G\' abor Szab\' o$^{1,6}$, \' Arp\' ad Moh\' acsi$^{1,6}$, Susanne Teschl$^4$,  Gerald Teschl$^5$,  
and  Julian King$^2$}
\address{$^1$MTA-SZTE Research Group on Photoacoustic Spectroscopy, D\'om t\'er 9, 6720 Szeged, Hungary}
 \address{$^2$Breath Research Institute, University of Innsbruck, Rathausplatz 4, A-6850 Dornbirn, Austria}
\address{$^{3}$University of Applied Sciences Vorarlberg, Hochschulstr.~1, A-6850 Dornbirn, Austria}
 \address{$^{4}$University of Applied Sciences Technikum Wien,  H\"ochst\"adtplatz 6, A-1200 Wien, Austria}
  \address{$^{5}$Faculty of Mathematics, University of Vienna, Oskar-Morgenstern-Platz 1, 1090 Wien, Austria}
    \address{$^{6}$Department of Optics and Quantum Electronics, University of Szeged, D\'om t\'er 9, 6720 Szeged, Hungary }
   \address{$^7$Univ.-Clinic for Anesthesia and Intensive Care, Innsbruck Medical University, Anichstr.~35, A-6020 Innsbruck, Austria}

\ead{aszabo@titan.physx.u-szeged.hu}
\ead{karl.unterkofler@fhv.at}
\ead{jking@oeaw.ac.at}

\begin{abstract}
 We develop a simple three compartment model based on mass balance equations which quantitatively  describes the  dynamics of breath methane concentration profiles during exercise on an ergometer. With the help of this model it is possible to estimate the endogenous production rate of methane in the large intestine by measuring breath gas concentrations of methane.
\end{abstract}

Dedicated  to  the memory of our friend, colleague, and mentor  Anton Amann.\\

\noindent{\it Keywords\/}: Modeling, Breath gas analysis,  Methane, Volatile organic compounds (VOCs), Production rates, SRI-PTR-TOF-MS, Exercise  \\[5mm]
Version: \today\\
J.\ Breath Res.\ {\bf 10}, 017105 (2016).

\maketitle

\section{Introduction} 
Various studies demonstrated that methane in humans majorly originates from anaerobic fermentation by methanogens in the large intestine. Methane can then traverse the intestinal mucosa and be absorbed into the systemic circulation. Since methane has a low solubility in blood, it is rapidly excreted by the lungs. It is a generally accepted criterion that a subject is considered to be a methane producer if the methane concentration in exhaled breath exceeds the ambient air level by 1~ppm \cite{bond1971, Costello2013}. Approximately 30-50\% of adults were found to be methane producers \cite{Triantafyllou2014}. Considering methane production, gender, age, and ethnic differences were observed \cite{Pitt1980,  Peled1985,  Levitt2006,   Polag2014}. Additionally, a significant day-to-day variation was reported \cite{Minocha1997}. However, the factors influencing the number of methanogens and the amount of methane produced are still unexplored.

The interaction between methanogens and gut function is an extensively studied field. Breath methane tests and culture based methods have traditionally been used to characterize methanogen populations \cite{Costello2013}. Culture based methods have high sensitivity; however they are cumbersome and time-consuming. Nevertheless the methane breath test is a convenient, quick and effective method for the assessment of methanogen populations; therefore it is increasingly used in the diagnostics of certain gastrointestinal conditions. In clinical practice, a combined hydrogen and methane breath test has been shown to be superior for the diagnosis of carbohydrate malabsorption syndromes and small intestinal bacterial overgrowth \cite{Costello2013}. It is commonly accepted that breath methane is associated with alterations in intestinal motility, and it is strictly related to constipation \cite{Roccarina2010, Furnari2012,  Gottlieb2015}. Additionally, numerous studies have found correlations between breath methane levels and diseases including colon-rectal cancer, irritable bowel syndrome and inflammatory bowel disease \cite{Roccarina2010, sahakian2010m, Kunkel2011, Furnari2012}. However, the results are controversial and the impact of endogenous bacterial methane generation on health is still not known with certainty.

Although numerous studies have conducted methane breath tests, there are only relatively few studies that investigated the routes of methane excretion, i.e., the correlation between methane concentration in breath and in the gut \cite{bond1971}. It is generally assumed that methane is not utilized by humans, and approximately 20\% of the methane produced by anaerobic fermentation is excreted by breath. The remaining 80\% is lost by flatus \cite{bond1971}.

It is worthwhile to note that a recent paper by Boros et al.\ reviewed the possible role of methane as a gasotransmitter \cite{boros2015}. It provides some evidence with respect to non-bacterial generation of methane in target cells which is possibly linked to mitochondrial dysfunction. Furthermore, methane-rich saline is hypothesized of having an anti-oxidative effect \cite{Chen20161}.

Breath tests can be performed even in real time allowing to monitor biological processes in the body. In our recent study the dynamics of endogenous methane release through the respiratory system have been investigated by measuring breath methane concentration profiles during exercise on an ergometer \cite{annaszabo1}. The qualitative behavior of such profiles was in good agreement with the Farhi equation~\cite{farhi1967} but the quantitative behavior deviated. The aim of this article is to develop a simple three compartment model to describe and explain quantitatively the observed breath methane concentration profiles. The present model can serve as a tool to estimate the endogenous production rate of methane in the large intestine from exhaled methane concentrations.

A list of symbols used is provided in Appendix~A.

\section{Measurements} \label{data}

\subsection{Setup}
End-tidal methane concentration profiles were obtained by means of a \emph{real-time} setup designed for synchronized measurements of exhaled breath VOCs as well as a number of respiratory and hemodynamic parameters. Our instrumentation has successfully been applied for gathering continuous data streams of these quantities during ergometer challenges~\cite{King2009} 
 as well as during sleep studies \cite{king2012a}. These investigations aimed at evaluating the impact of breathing patterns, cardiac output or blood pressure on the observed breath concentration and permit a thorough study of characteristic changes in VOC output following variations in ventilation or perfusion. 
 An extensive description of the technical details is given in a previous work \cite{King2009}.

\begin{figure}
\centering
\begin{tabular}{c}
\hspace{-6mm} \includegraphics[width=9cm]{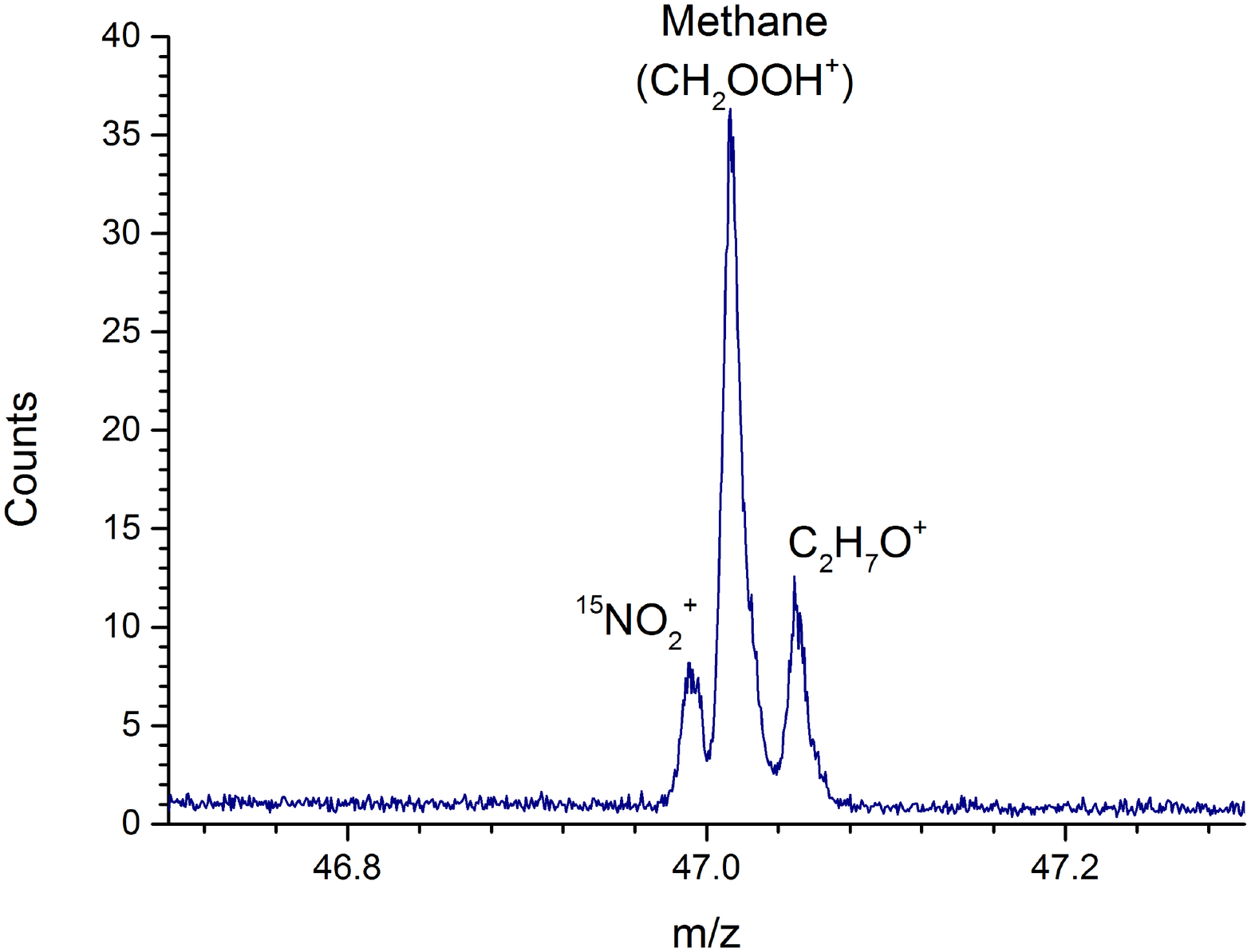}
\end{tabular}
\caption{Spectrum of methane as measured by SRI-PTR-TOF-MS using $\mathrm{O}_{2}^{+}$ primary ions.}\label{fig:methaneTOF}
\end{figure}

In brief, the core of the mentioned setup consists of a head mask spirometer system allowing for the standardized extraction of arbitrary exhalation segments, which subsequently are directed into a 
Selective Reagent Ionization Proton Transfer Reaction Time of Flight Mass Spectrometer
(SRI-PTR-TOF-MS, Ionicon Analytik GmbH, Innsbruck, Austria) for online analysis.
This analytical technique has proven to be a sensitive method for the quantification of volatile molecular species $M$ down to the ppb (parts per billion) 
range. To measure methane we took advantage of the reaction of the primary $\mathrm{O}_{2}^{+}$ precursor with methane \cite{Adams1980, Dryahina2010, Wilson2003} 
\begin{eqnarray}
\mathrm{O}_{2}^{+} + \mathrm{C}\, \mathrm{H}_{4} \to \mathrm{C}\, \mathrm{H}_{2}\,\mathrm{O} \mathrm{O} \mathrm{H}^{+} + \mathrm{H} \nonumber .
\end{eqnarray}
Count rates of the resulting product ion appear at the specified mass-to-charge ratio $m/z=47.0128$ 
(see Figure~\ref{fig:methaneTOF} and figure~6 in \cite{kingIWISH2012}) and can subsequently be converted to absolute concentrations by means of calibrations factors obtained from analyzing calibrations mixtures containing a known amount of methane and humidity. 

\begin{figure}[h]
\centering
\begin{tabular}{c}
\includegraphics[height=83mm]{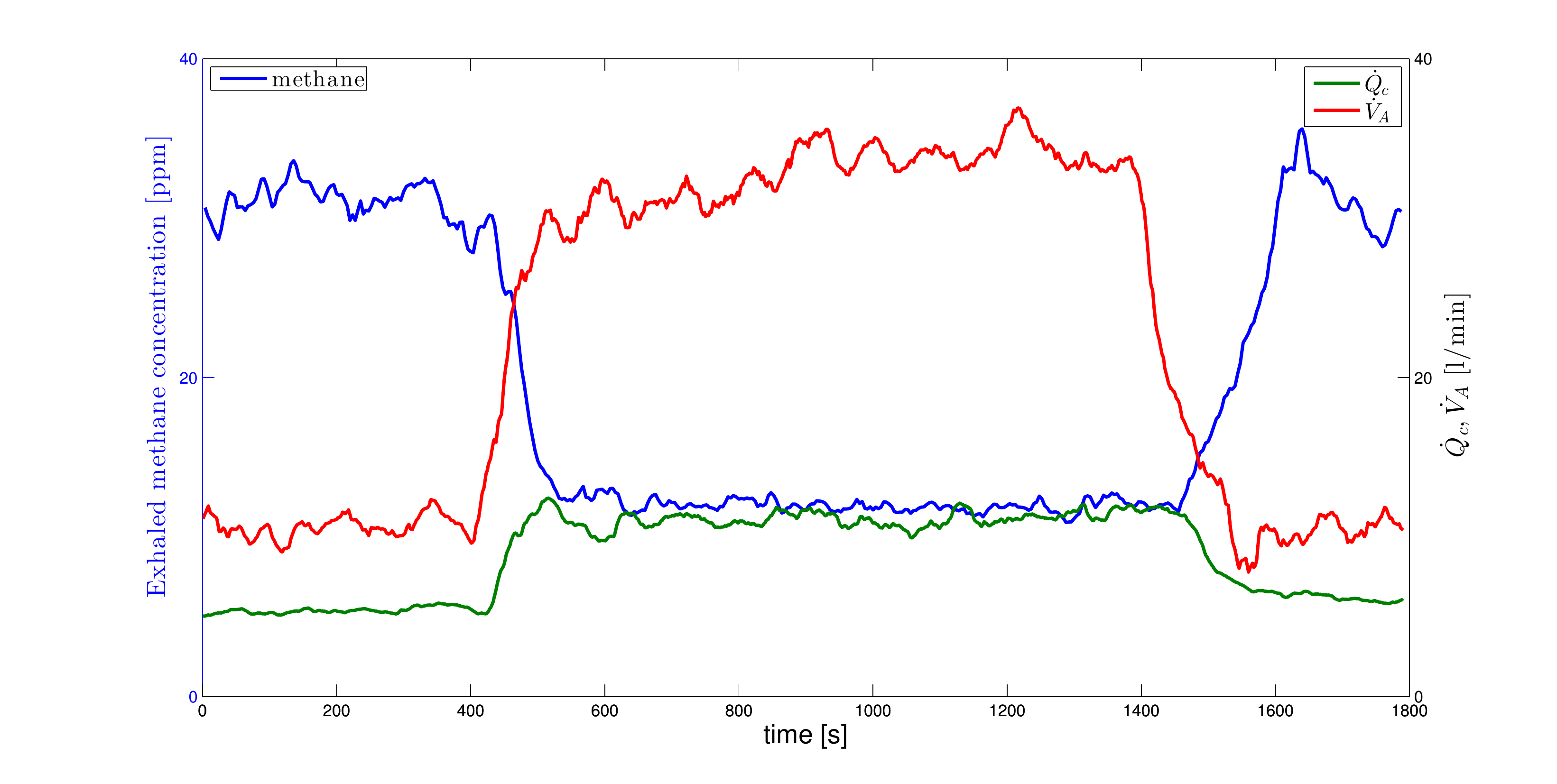}
\end{tabular}
\caption{Typical result of  an ergometer session for one single volunteer. Average values: cardiac output (green), 
rest: $\qc=5.41$ [$\ell$/min]; $75$~Watts: $\qc=11.07$ [$\ell$/min];
alveolar ventilation (red), rest: $\qalv=10.69$ [$\ell$/min]; $75$~Watts: $\qalv=33.12$ [$\ell$/min];
and exhaled end-tidal (nose sampling) methane levels  (blue), rest: $\calv = 31.08 $ [ppm]; $75$~Watts: $\calv = 11.92$ [ppm],
room air concentration of methane:  $3.37$ [ppm].}\label{fig:methaneKAUN}
\vspace{2mm}
\end{figure}

So far, only some preliminary measurements were carried out by means of the setup described above. Two healthy 
methane producing adult volunteers (one male, one female) were asked to perform several ergometer challenges of
approximately 6 minutes rest, 17 minutes with 75 Watts, and then approximately 6 minutes rest again.
The exact protocol was:
\begin{itemize}
\item{} seconds 0--380: the volunteer rests on the ergometer
\item seconds 380--1400: the volunteer pedals at a constant workload of 75 Watts
\item seconds 1400--1800: the volunteer rests on the ergometer
 
\end{itemize}
Figure~\ref{fig:methaneKAUN} shows a tyical result of such an ergometer session for one volunteer. While the number of probands is certainly very limited, the relative changes of breath methane concentrations are in good agreement with similar measurements employing a different analytical set up   as described in a recent work  \cite{annaszabo1} (see figure~1 therein).\\

 \section{Modeling  methane distribution in the body}

\subsection{Methane exchange in the lungs} 
In humans, methane is mainly produced by enteric bacteria in the large intestine and distributed within the body by the venous blood leaving the intestine. When reaching the lungs, it is partially released into breath.
The amount of methane transported at time $t$ to and from the lungs via  blood flow is given by
\begin{eqnarray}
\qc(t)\big(\cven(t)-\cart(t)\big), \nonumber
\end{eqnarray}
where $\qc$ denotes the cardiac output,  $\cven$  the averaged mixed venous concentration,
  and $\cart$ is the arterial concentration.
  
On the other hand one in- and exhales the amount
\begin{eqnarray}
\qalv(t)\big(\cinh-\calv(t)\big), \nonumber
\end{eqnarray}
where $\qalv$ denotes the alveolar ventilation,  $\cinh$ denotes the concentration in inhaled air, and $\calv$ the alveolar concentration.
While $\cinh$ is assumed to be zero for many endogenous VOCs, the current  average atmospheric methane concentration level is about 1.8~[ppm]  \cite{Nisbet2014} and can hence not be neglected\footnote{Typical room air concentrations are often even higher than 1.8 ppm.}.

Combining these two terms leads to the following mass balance equation for the lungs\footnote{For notational convenience we have dropped the time variable  $t$, i.e., we  write $C_{X}$ instead of $C_{X}(t)$, etc. $C_{X}$ denotes the instant or averaged concentration of $X$ over a small sampling period 
$\tau$, i.e.,  
 $C_{X}(t) = 1/ \tau \, \int_{t-\tau/2}^{t+\tau/2} C_{X}(s) ds$.} 
\begin{eqnarray}  \label{eq:alv}
\valv\frac{d\calv}{dt}=\qalv(\cinh-\calv)+\qc(\cven-\cart),
\end{eqnarray}
where $\valv$ denotes the  volume of the lung. Both sides of Equation~(\ref{eq:alv}) have units $\mu$mol/min (compare Appendix~A).

If the system is in an equilibrium state (e.g., stationary at rest) Equation~(\ref{eq:alv}) 
reads $0   =   \qalv\big(\cinh-\calv(\cinh) \big)+\qc\big(\cven(\cinh)-\cart\big)$ and 
using Henry's law 
\begin{equation} \label{henryslaw}
 \cart= \hen \,\calv  
\end{equation}
we obtain
\begin{eqnarray}
\calv(\cinh)& = & \frac{  \cinh }{\frac{\hen}{r} + 1}+ \frac{  \cven(\cinh)}{\hen + r} \label{FarhiCinh}
\end{eqnarray}
where $r=\dot{V}_{A}/\dot{Q}_{c}$ is the ventilation-perfusion ratio and $\hen$ denotes the blood:air partition coefficient.

{Remark:} The modeling approach followed above is only valid for VOCs with blood:air partition coefficient  less than $10$, i.e., compounds for which the upper airways have no influence on the observable breath concentrations \cite{anderson2003}. Methane with a
blood:air partition coefficient $\hen= 0.066$ \cite{poyart1976} fulfills this requirement.

Since  $\hen$ for methane is so small 
we get  $\cven(\cinh) \approx \cven(0) $,  $ 1+\frac{\hen}{r} \approx 1 $, and $\hen + r\approx r $. From this follows that for  methane it suffices to subtract the inhaled methane concentration to correct for room air levels  (see a previous work for more details  \cite{unterkofler2015}).

Thus Equation~(\ref{FarhiCinh})  can be simplified to

\begin{equation}
\calv(0)=\calv(\cinh)-\cinh =    \frac{\qc}{\qalv}\, \cven(0) = \frac{1}{r}\,  \cven(0) . \label{eq4}
\end{equation}

When a subject is under  constant conditions at rest, 
 $\cven$ is approximately constant. From Equation~(\ref{eq4})  it may then be concluded that variations in the  alveolar concentration $\calv(0)=(\calv(\cinh)-\cinh)$ directly reflect changes in ventilation (e.g., due to altered breathing frequency) and perfusion (e.g., due to altered heart rate). This can be tested by forced hypo- and hyperventilation at rest    as shown in figure~5 in a previous work   \cite{annaszabo1}.

The ventilation-perfusion ratio $r$ 
 is approximately one at rest but substantially increases for a moderate exercise regime at $75$ Watts, 
since the cardiac output increases approximately two-fold while the ventilation increases three- to four-fold~\cite{King2009}.
Consequently, one would expect from Equation~(\ref{eq4}) that the alveolar methane concentration should decrease by a factor of approximately 1.5--2 when exercising at that workload.   

Contrary to this prediction, measurements of breath methane concentrations show a drop by a factor of 3 to 4 when exercising at $75$ Watts \cite{kingIWISH2012, annaszabo1}, see also Figure~\ref{fig:methaneKAUN}.

\subsection{A three compartment model} \label{rationale}

The intuitive rationale for this phenomenon is as follows. The intestinal bacteria are the main source of methane. 
At rest, the intestine receives about 15\% of the  total blood flow of approximately $5$~$\ell$/min, leading to an absolute perfusion of approximately $0.75$~$\ell$/min, which is matched to the metabolic needs of gut tissue.
When exercising moderately, this absolute blood flow to the intestine may be assumed constant, since its metabolic needs remain largely unchanged. However, the relative (fractional) blood flow to the intestine decreases, as a major part of total cardiac output is now directed to the working muscles. As a result, the relative contribution of intestinal venous blood (characterized by high methane concentrations) to mixed
venous blood will be reduced, causing the mixed venous methane concentration to drop. The decrease in breath methane concentrations during exercise may hence be interpreted as a combination of two separate effects: an increased dilution within the lungs due to an increased ventilation-perfusion-ratio (cf.~Equation~(\ref{eq4})) and an additional reduction of the mixed venous concentration levels due to a reduced fractional perfusion of the intestine.

\begin{figure}[ht]

\centering
\centering
\begin{picture}(11,7.8)

\put(8,6.3){lung compartment}
\put(8,3.3){\parbox{3cm}{richly perfused\\ compartment}}
\put(8,0.9){gut compartment}

\put(1,6.8){\line(0,1){1}}
\put(1.7,7.5){\line(0,1){0.3}}
\put(7,6.8){\line(0,1){1}}
\put(6.3,7.5){\line(0,1){0.3}}

\put(1.7,7.5){\line(1,0){4.6}}
\put(1,6.8){\line(1,0){2}}\put(5,6.8){\line(1,0){2}}
\put(3,6.8){\line(0,-1){0.6}}
\put(5,6.8){\line(0,-1){0.6}}
\put(1,6.2){\line(1,0){2}}\put(5,6.2){\line(1,0){2}}
\put(1.7,5.5){\line(1,0){4.6}}

\multiput(3.05,6.5)(0.2,0){10}{\line(1,0){0.1}}

\put(2,7){$\cinh$}
\put(2,5.7){$\cven$}

\put(5.6,7){$\calv$}
\put(5.6,5.7){$\cart$}

\put(2.5,7.1){\vector(1,0){3}}
\put(3.8,7.2){$\scriptstyle\qalv$}
\put(1.35,7.1){\line(1,0){0.5}}
\put(1.35,7.7){\vector(0,-1){0.3}}
\put(1.35,7.4){\line(0,-1){0.3}}
\put(6.15,7.1){\line(1,0){0.5}}
\put(6.65,7.1){\vector(0,1){0.6}}

\put(2.5,5.9){\vector(1,0){3}}
\put(3.8,6){$\scriptstyle\qc$}

\put(1,6.2){\line(0,-1){4.9}}
\put(7,6.2){\line(0,-1){4.9}}

\put(1.7,5.5){\line(0,-1){1}}
\put(6.3,5.5){\line(0,-1){1}}

\put(1.7,4.5){\line(1,0){4.6}}
\put(1.7,3.8){\line(1,0){1.3}}\put(5,3.8){\line(1,0){1.3}}
\put(3,3.8){\line(0,-1){1.3}}
\put(5,3.8){\line(0,-1){1.3}}
\put(3,2.5){\line(1,0){2}}

\multiput(3.05,3.5)(0.2,0){10}{\line(1,0){0.1}}

\put(5.5,4.1){\vector(-1,0){2.8}}
\put(3.5,4.2){$\scriptstyle(1-\qper)\qc$}

\put(5.6,4){$\cart$}
\put(1.8,4){$\crptb$}
\put(3.7,2.9){$\crpt$}
\put(5,3.25){\vector(1,0){0.5}}
\put(5.6,3.15){$\ml$}
\put(5.5,2.75){\vector(-1,0){0.5}}
\put(5.6,2.65){$\prl$}

\put(1.7,3.8){\line(0,-1){1.8}}
\put(6.3,3.8){\line(0,-1){1.8}}

\put(1.7,2){\line(1,0){4.6}}
\put(1,1.3){\line(1,0){2}}\put(5,1.3){\line(1,0){2}}
\put(3,1.3){\line(0,-1){1.3}}
\put(5,1.3){\line(0,-1){1.3}}
\put(3,0){\line(1,0){2}}

\multiput(3.05,1)(0.2,0){10}{\line(1,0){0.1}}

\put(5.5,1.6){\vector(-1,0){2.8}}
\put(3.5,1.7){$\scriptstyle\qper\qc$}

\put(5.6,1.5){$\cart$}
\put(1.8,1.5){$\cperb$}
\put(3.7,0.4){$\cper$}
\put(5.5,0.25){\vector(-1,0){0.5}}
\put(5.6,0.15){$\prm$}

\put(1.35,1.6){\line(1,0){0.3}}
\put(1.35,1.6){\vector(0,1){1.5}}
\put(1.35,4.1){\line(1,0){0.3}}
\put(1.35,4.1){\circle*{0.1}}
\put(1.35,3.1){\vector(0,1){2}}
\put(1.35,5.9){\line(1,0){0.5}}
\put(1.35,5.1){\line(0,1){0.8}}

\put(6.65,1.6){\line(-1,0){0.5}}
\put(6.65,3.1){\line(0,-1){1.5}}
\put(6.65,4.1){\line(-1,0){0.5}}
\put(6.65,4.1){\circle*{0.1}}
\put(6.65,5.1){\vector(0,-1){2.3}}
\put(6.65,5.9){\line(-1,0){0.5}}
\put(6.65,5.9){\vector(0,-1){1}}

\end{picture}
\caption{Three compartment model for methane: lung compartment with gas exchange, gut compartment with production of methane by enteric bacteria, and richly perfused tissue compartment containing the rest of the body including muscles (possible but small production and metabolic rate)}\label{fig:model_struct}

\end{figure}
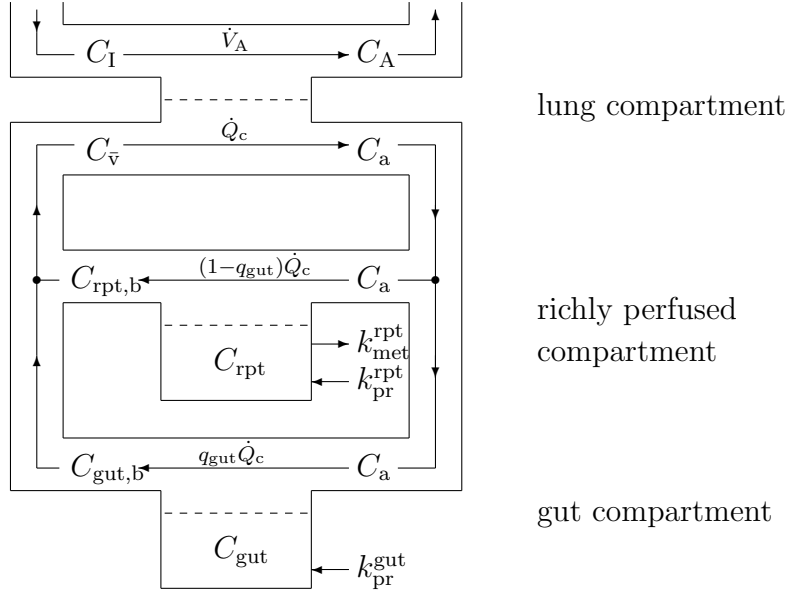

In order to mathematically capture the mechanism illustrated above, we developed a three compartment model based on mass balance equations, similar to previous modeling efforts, e.g., with respect to isoprene~\cite{King:isoprene}.
The model consists of a lung compartment, a  gut compartment (intestine), and a richly perfused compartment which comprises the rest of the body as shown in Figure~\ref{fig:model_struct}.

The mass balance equation for the lung compartment has already been derived in Equation (\ref{eq:alv}).
Arterial blood leaving the heart with concentration $\cart$ is divided into two blood flows
 $\qper\qc$ and $(1-\qper)\qc$, where $\qper$ denotes the fractional blood flow to the intestine.
 
 The molar flow to and from the gut compartment is given by $\qper\qc\cart$ and $\qper\qc\lper\cper$, respectively, where the proportional factor $\lper$ is the corresponding blood:tissue partition coefficient. This yields the following mass balance equation for the gut compartment (intestine):
 \begin{equation}\label{eq:per}
\vper\frac{d\cper}{dt}=\qper\qc(\cart -\lper\cper)+ \mu\, \prm . 
\end{equation}
Here, $\vper$ denotes the effective volume\footnote{The vascular blood compartment and the intracellular tissue compartment are assumed to be in an equilibrium and therefore can  be combined into one single gut compartment with an effective volume.  For more details about effective volumes compare appendix~2  in a previous paper \cite{king2010a}.} of the gut. The factor $\mu \, \approx 0.2$  respects the fact  that $80$\% of methane is lost by flatus and therefore does not enter the blood stream~\cite{bond1971}.
In addition we assume that within the time frame of the ergometer sessions presented, the net production rate $\prm$  of methane stays constant and a possible metabolization in the large intestine can be  respected by a  correction of $\prm $.
Both sides of Equation~(\ref{eq:per}) have units $\mu$mol/min (compare Appendix~A).

Analogously, for the richly perfused tissue compartment containing the rest of the body including muscles we get
\begin{eqnarray} \label{eq:rpt}
\vrpt\frac{d\crpt}{dt}=(1-\qper)\qc(\cart -\lrpt\crpt)-\ml \lrpt \crpt + \prl,
\end{eqnarray} 
where $\vrpt$ denotes the effective volume of this compartment, $\prl$ respects a possible small nonbacterial production rate and $\ml$ represents a possible small metabolic rate\footnote{Here we used the usual convention to multiply $\ml$ by $ \lrpt$. It would be more natural to use $\ml$ only.}.
Both sides of Equation~(\ref{eq:rpt}) have units $\mu$mol/min (compare Appendix~A).\\
{ Remark:} According to Bond~\cite{bond1971} both $\ml$ and $\prl$ are  very small and hence can be neglected in a first modeling approach.  

The mixed venous concentration is given by the weighted sum of the two body compartment concentrations
\begin{equation}\label{eq:cven}
\cven:=(1-\qper)\lrpt\crpt+\qper\lper\cper.
\end{equation}
The total mass balance given by the Equations (\ref{eq:alv}), (\ref{eq:per}), and (\ref{eq:rpt}) constitutes a coupled system of three first order ordinary differential equations (ODEs) of the form
\begin{equation} \label{eq:system}
\frac{d}{dt} \vc(t) = \vg(t,\vc(t)) =: A(t) \, \vc(t) + \vb(t)
\end{equation}
for the three unknown concentrations
\begin{equation}
\vc(t) = \left( \cart(t), \crpt(t), \cper(t) \right).
\end{equation}
The matrix $A(t)$ and the vector $\vb(t)$ are given by 
\begin{eqnarray}
\hspace{-15mm} A(t)=
\left( \begin{array}{ccc}
- \frac{\qalv(t)+ \hen \qc(t)}{\valv} & (1- \qper(t)) \hen \lrpt  \frac{ \qc(t)}{ \valv} & \qper(t) \hen \lper  \frac{ \qc(t)}{ \valv} \\
(1-\qper(t)) \frac{ \qc(t)}{ \vrpt}  & -  \frac{(1-\qper(t)) \lrpt \qc(t)+ \lrpt \ml}{ \vrpt}  & 0 \\
\qper(t) \frac{ \qc(t)}{ \vper} & 0 & -\qper(t) \frac{ \qc(t)}{ \vper} \lper
\end{array} \right)  \nonumber ,
\end{eqnarray}

\begin{eqnarray}
  \vb(t)=
\left( \begin{array}{c}
\hen \frac{ \qalv(t)}{\valv} \cinh \\
 \frac{ \prl}{ \vrpt} \\
\mu\frac{ \prm}{ \vper}
\end{array} \right) . 
\end{eqnarray}

All external inputs $(\qalv(t), \qc(t),   \cinh)$ affecting the system can be measured by means of the experimental setup and are therefore assumed to be known.
The partition coefficients $\hen, \lrpt, \lper$ may partially be derived from literature values, see Section~\ref{simulation}.

Model parameters that are a priori unknown and not directly measurable include the metabolic rate $\ml$,  the production rates $\prm$ and $\prl$, as well as  
 the effective volumes $\vrpt$, $\vper$, $\valv$, which influence the time constants for achieving a steady state.  These will either have to be fixed at best-guess values or estimated from the measurement data by means of a suitable parameter estimation scheme, see Section~\ref{simulation}.

As explained in the model rationale, the absolute blood flow through the intestine is postulated to stay approximately constant during moderate exercise. We therefore use the following simple model for the fractional blood flow $\qper$
\begin{equation}\label{eq:qgut}
\qper(t) = q_0\,  \frac{\dot{Q}_{c,rest}}{\qc(t)}, \qquad q_0 = 0.15
\end{equation}
where $\dot{Q}_{c,rest}  $ is the average total blood flow (cardiac output) at rest.

In addition, the methane concentration in exhaled end-tidal  air  is measured and
 assumed to be equal to the alveolar concentration
\begin{equation}\label{eq:meas}
y(t):=\cmeas(t)=\calv(t) =\hen^{-1} \cart(t).
\end{equation}

\subsection{Steady state analysis} \label{stationary}
When in a steady state the system of differential equations reduces to the following simple linear algebraic system
\begin{eqnarray}
0 & = &\qalv\big(\cinh-\calv\big)+\qc\big(\cven-\cart\big), \nonumber\\
0 & = &\qper\qc(\cart -\lper\cper)+\mu \prm, \nonumber\\
0 &= &(1-\qper)\qc(\cart -\lrpt\crpt)-\ml\lrpt\crpt +\prl .
\end{eqnarray}
Solving with respect to $\cper,\crpt$, and $\prm$ yields
\begin{eqnarray}
\crpt & =& \frac{\hen}{\lrpt}\, \frac{(1-\qper) \qc \,\calv+\prl}{(1-\qper) \qc+\ml}  , \nonumber\\
\cper & =& \frac{\hen}{\lper}\, \frac{\calv+ \frac{r}{\hen}(\calv-C_{I})}{\qper} - 
\frac{(1-\qper)}{\lper\, \qper}\, \frac{\hen (1-\qper) \qc \, \calv + \prl}{(1-\qper) \qc + \ml}, 
\nonumber \\
\prm & =& \frac{1}{\mu} \Big(\dot{V}_{A}  (C_{A}-C_{I}) + \frac{(1-\qper) \qc}{(1-\qper) \qc+\ml}    (\hen \, \ml \,\calv- \prl ) \Big) .\nonumber\\ \label{eq:stationary}
\end{eqnarray}

If we assume the nonbacterial production and the metabolic rate  in the richly perfused compartment to be negligible we set $ \prl=0$ and $\ml = 0$, respectively. Then Equation~(\ref{eq:stationary})  simplifies to
\begin{eqnarray} \label{stationaryEqn}
\crpt(\cinh) & = &\frac{\hen}{\lrpt}\, \calv(\cinh) ,\nonumber \\
\cper(\cinh) & =&  \frac{\hen}{\lper}\, \, \calv(\cinh) + \frac{r}{\qper\, \lper}(\calv(\cinh)-\cinh)  , \nonumber \\
\prm & = &\frac{1}{\mu} \dot{V}_{A}  (\calv(\cinh)-\cinh) .
\end{eqnarray} 
Furthermore, we recall that
\begin{eqnarray}
\cven(\cinh)  &=&  (1-\qper)\hen\, \calv(\cinh)+\qper\lper\, \cper(\cinh), \nonumber \\  \cart(\cinh)  &=& \hen \,\calv(\cinh)  . \nonumber
\end{eqnarray}
Here we explicitely indicated the dependence of the various quantities on the inhaled concentration $\cinh$.

From Equation~(\ref{stationaryEqn}) we conclude that for a steady state (e.g., at rest or at a moderate constant workload):
\begin{enumerate}
\item[(i)]
 The methane concentration $\crpt(\cinh)$ in the richly perfused tissue compartment is proportional to the alveolar concentration. However, $\crpt(\cinh)$ is much smaller than $\calv(\cinh)$ since $\hen$ is very small.
\item[(ii)]Analogously, since $\hen$ is small for $\cper$ we get
\begin{eqnarray}
\hspace{-15mm} \cper(\cinh)  \approx \frac{r}{\qper\, \lper}(\calv(\cinh)-\cinh) =   \frac{r}{\qper\, \lper}\, \calv(0) \approx \cper(0) \nonumber
\end{eqnarray}
or, vice versa, 
\begin{eqnarray}\label{stationaryEqn2}
\hspace{-15mm} \calv(0) &  \approx  \frac{\qper\, \lper}{r}\, \cper(0), 
\end{eqnarray} 
showing that the breath methane concentration is roughly proportional to the fractional intestinal blood flow $\qper$.

\item[(iii)] Since we expect  $\prm$ to be constant on a ``medium time scale''  
(e.g., during an ergometer session) we  obtain 
\begin{eqnarray}
\calv(0)= \calv(\cinh) - \cinh = \mu\, \prm \, \frac{1}{\qalv }.
\end{eqnarray} 

Thus the product $\calv(0) \times \qalv$ 
does not change when switching from one stationary regime to another, e.g. when switching from a resting steady state to an exercise steady state at 75~W, viz.,
\begin{eqnarray}  
\frac{1}{\mu}\, \dot V_{\mathrm{A},rest} \, C_{\mathrm{A},rest}(0)= \prm  =
 \frac{1}{\mu}\, \dot V_{\mathrm{A},75Watts} \, C_{\mathrm{A},75Watts}(0) .
\end{eqnarray}

This explains the  experimental findings of a recent work \cite{annaszabo1} (see figure 3 therein).
 The production rate of methane in the intestine can therefore be estimated by taking  the product of  average steady state values of $\qalv$  and $\calv(0)$,
\begin{eqnarray}  
 \prm & = \frac{1}{\mu}\, \qalva \, \calva(0) . \label{eqn:prodshort}
\end{eqnarray}

\end{enumerate}

\subsection{  Simulation of an ergometer session and parameter estimation} \label{simulation}

In this section we calibrate 
the proposed model based on the physiological data presented in Figure~\ref{fig:methaneKAUN}, corresponding to one \emph{single} representative volunteer   
following the line of a previous work~\cite{King:isoprene}.
 It will turn out that the model appears to be flexible enough to capture the methane profiles in exhaled breath generally 
 observed during moderate workload ergometer challenges as conducted in a recent work~\cite{annaszabo1}. 
 In a first attempt we  set the parameter describing   
 a possible small nonbacterial production rate to zero, i.e., we fix $\prl=0$.
For $ \vrpt$ and $ \valv $ we use the nominal values   $\valv =4.1$~[$\ell$] and $\vrpt=15.22$~[$\ell$] (compare with table C.1 1 in a previous work \cite{King:isoprene}), and $\vper=1$~[$\ell$].

The remaining unspecified parameters $p_j \in \{ \prm,  \ml\} $ may be estimated from the knowledge of measured breath methane concentrations $y$ by means of parameter estimation.
More specifically, the \emph{subject-dependent} parameter vector 
\[\tb=( \prm,  \ml )\]
as well as the nominal  steady state levels $\cb_0=\cb(t_0)$ can be extracted by solving the ordinary least squares problem
\begin{equation}\label{eq:ls}
\textunderset{$\tb$, $\cb_0$}{argmin}
\sum_{i=0}^n \big(y_i-\calv(t_i)\big)^2,
\end{equation}
subject to the constraints
\begin{equation}\label{eq:const}\left\{\begin{array}{ll}
\gb(t_0,\cb_0,\tb)=\nulb & \textrm{(steady state)}\\
\tb,\cb_0 \geq \nulb & \textrm{(positivity).}
     \end{array} \right.
\end{equation}
Here, $\gb$ is the right-hand side of the ODE system~(\ref{eq:system}),   and $y_i=C_{\mathrm{meas},i}$ is the measured end-tidal methane concentration at time instant $t_i$ ($t_0=0$). 

For this purpose the measured physiological functions $\qalv$ and $\qc$ were converted to input function handles by applying a local smoothing procedure to the associated data and interpolating the resulting profiles with splines. 
Furthermore, while the richly perfused compartment so far has been treated as an abstract control volume without particular reference to any specific tissue group, for identifiability reasons we now set $\lrpt=1$ as well as $\lper=1$
which corresponds to the in vitro blood:tissue methane partition coefficient for brain tissue in rabbits~\cite{ohta1979}, as currently no further values have been published. Initial concentrations and fitted parameters are given in Table~\ref{table:fit}.

\begin{table}[H]
\centering \footnotesize
\begin{tabular}{|lcc|}\hline
{\large\strut}Variable&Symbol&Fitted value (units)  \\
 \hline \hline 
{\large\strut}Production intestine &$\prm$ & $51.4$ [$\mu$mol/min] \\
{\large\strut}Metabolic rate  &$\ml$ & $0.01$ [$\ell$/min] \\
{\large\strut}Initial concentration alveoli $(t=0)$ &$\calv$ & 1.15 [$\mu$mol/$\ell$]  \\
{\large\strut}Initial concentration rpt $(t=0)$ &$\crpt$ & 0.076 [$\mu$mol/$\ell$]  \\
{\large\strut}Initial concentration intestine  $(t=0)$&$\cper$ & 13.6 [$\mu$mol/$\ell$]  \\
\hline
\end{tabular}
\caption[small]{
Decisive model parameters resulting from the fit in Fig.~\ref{fig:fit}. 
}
\label{table:fit}
\end{table}

All estimated quantities for the test subject under scrutiny take values in a physiologically plausible range. According to Equations~(\ref{henryslaw}) and~(\ref{eq:cven}), arterial and mixed venous blood concentrations at the start of the experiment are estimated for $t=0$ as $\cart =0.076$~$\mu$mol/$\ell$ and $\cven =2.1$~$\mu$mol/$\ell$, respectively.
Total endogenous production is estimated to equal approximately  $51.4$~[$\mu$mol/min].
The simulation indicates also a very small metabolic rate of $0.01$~[$\ell$/min], which is negligible compared to the production rate.

\begin{figure}
\centering
\begin{tabular}{c}
\hspace{-15mm} \includegraphics[width=10cm]{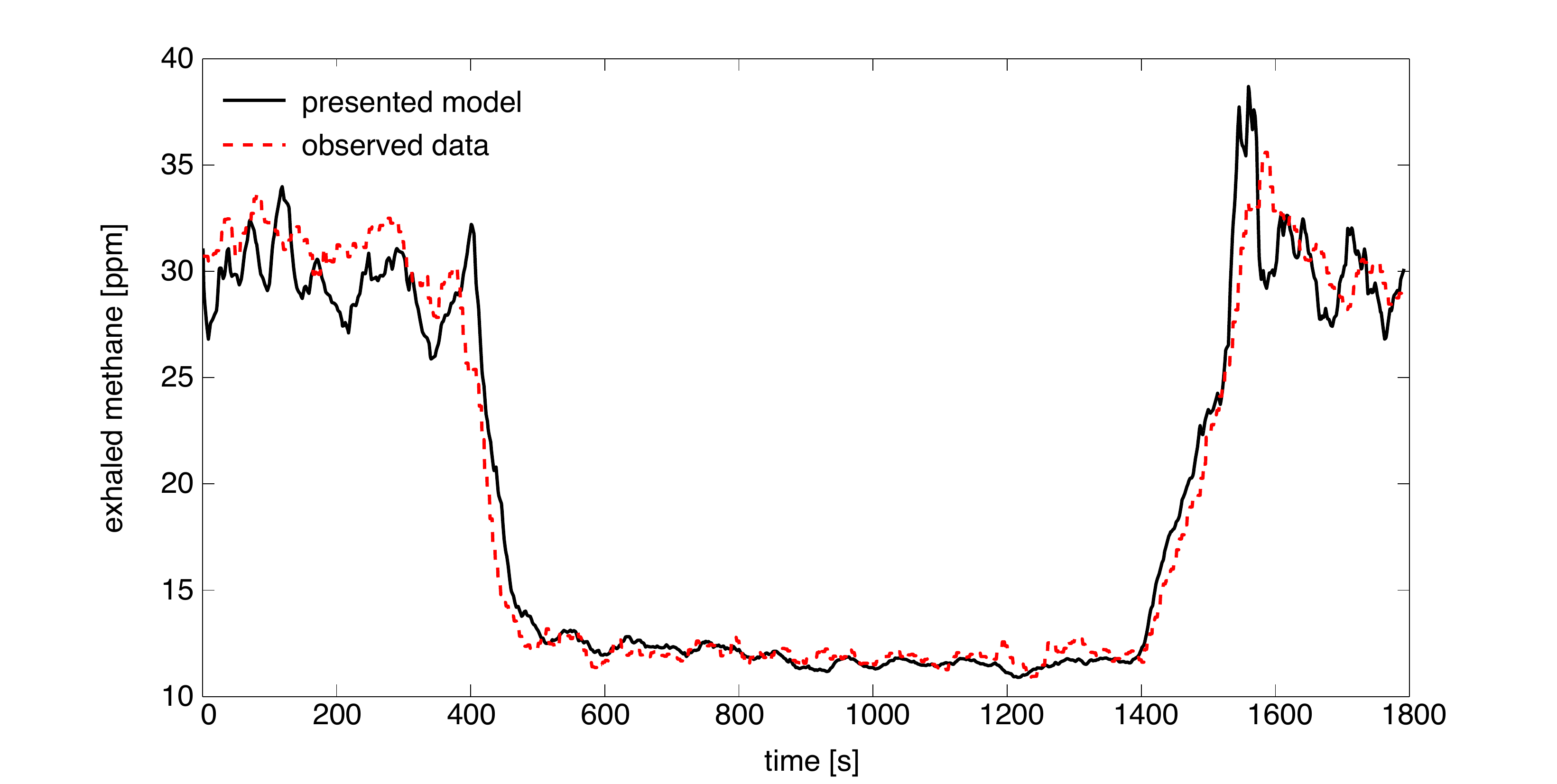}\\
\hspace{-15mm} \includegraphics[width=10cm]{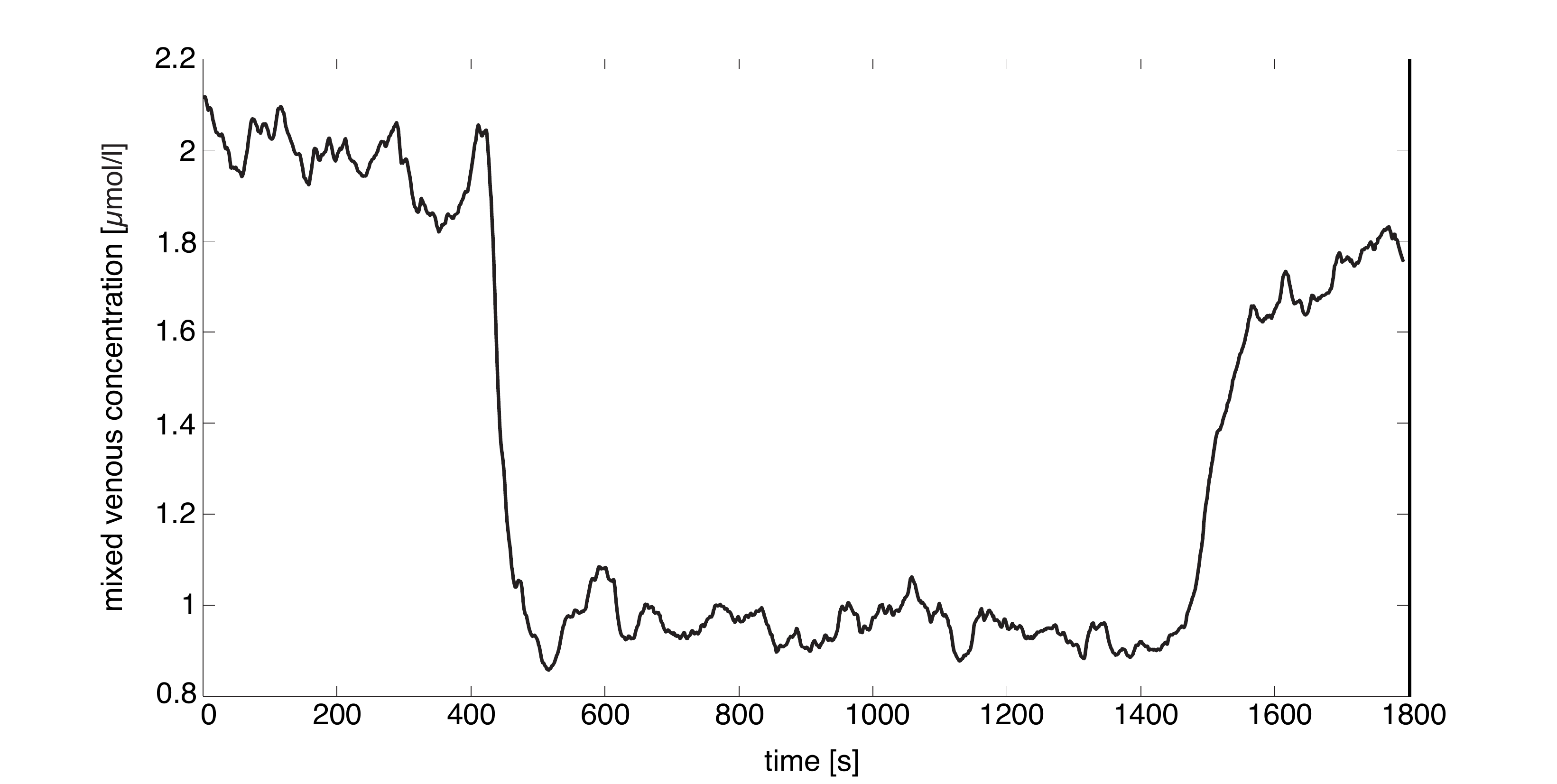}\\
\hspace{-15mm} \includegraphics[width=10cm]{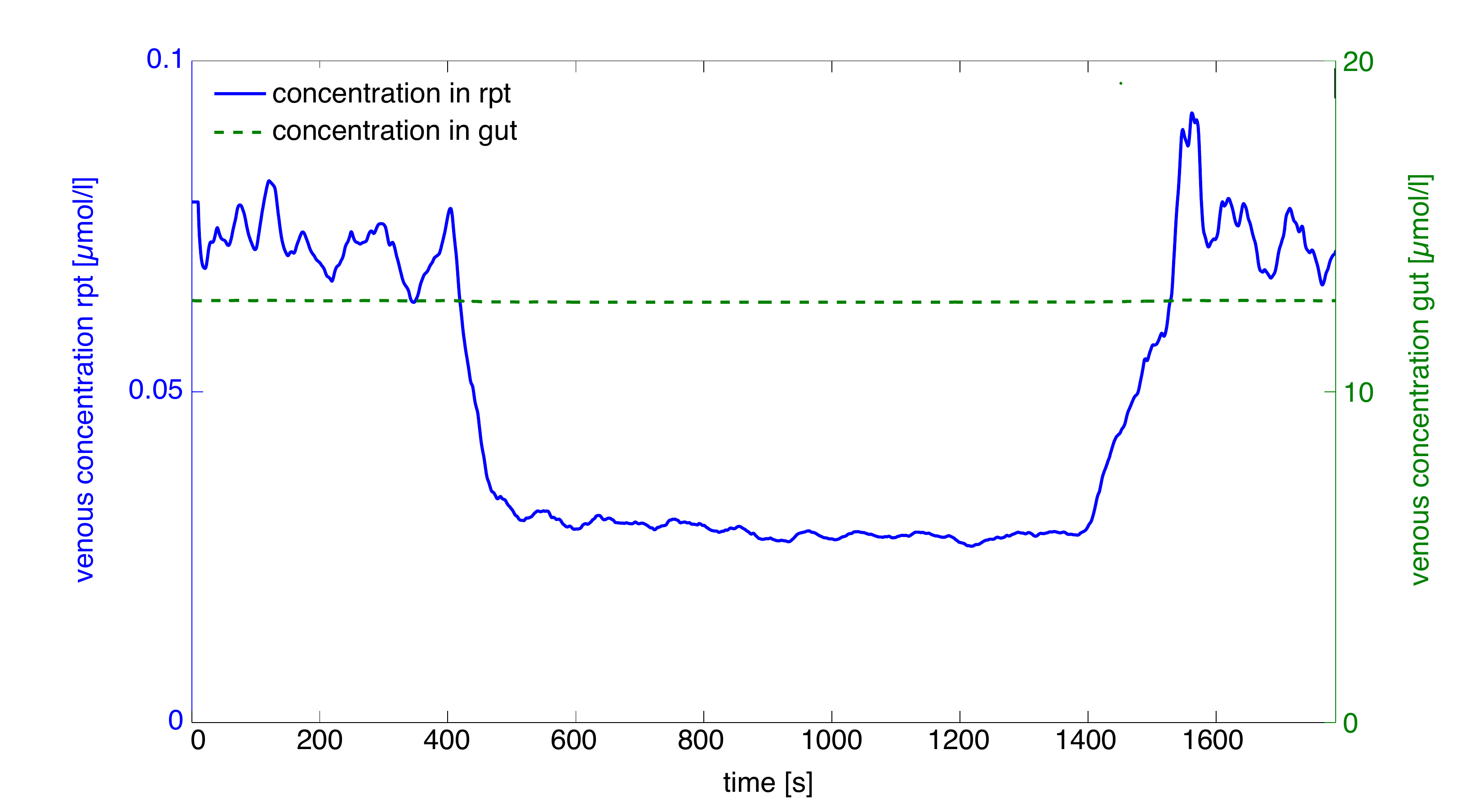}\\
\hspace{-15mm} \includegraphics[width=10cm]{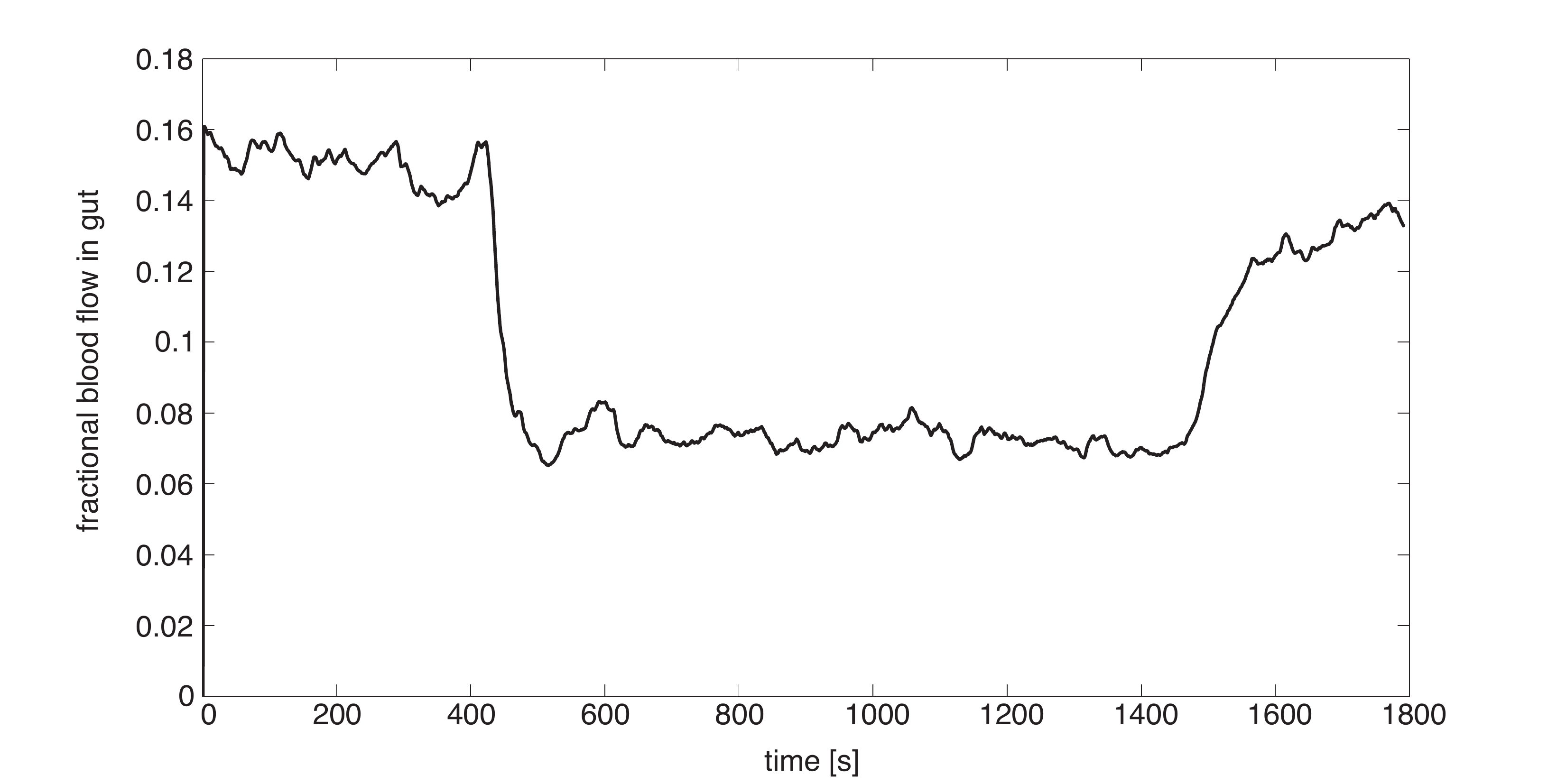}
\end{tabular}
\caption{
First panel: simulation of end-tidal methane concentration behavior during exercise conditions, cf.~Fig.~\ref{fig:methaneKAUN}.
Second panel: predicted methane concentrations in mixed venous blood ($\cven$).
Third panel: venous blood concentration returning from the gut ($\cper$) and returning from the richly perfused tissue ($\crpt$). 
Fourth panel: predicted profile of the fractional  gut blood flow 
$\qper$ according to Equation~(\ref{eq:qgut}).}\label{fig:fit}
\end{figure}

The results of the simulation are presented in  Fig.~\ref{fig:fit}. The first panel of Figure~\ref{fig:fit} shows that the methane concentration profiles obtained from the experiment and from the model are in good agreement. This suggests that the three-compartment model can describe quantitatively the methane profile changes during an ergometer challenge, while the  Farhi equation provided solely qualitative agreement \cite{annaszabo1}.

\section{Conclusion}
Despite the fact that methane breath tests are now widely accepted in clinical practice, a quantitative description of the routes of methane excretion is still lacking. The present paper intends to fill this gap by introducing a model for the distribution of methane in various parts of the human body. Particularly, we aimed at capturing the exhalation kinetics of breath methane in response to exercise. Classical pulmonary inert gas elimination theory  according to the Farhi equation~\cite{farhi1967} is deficient in this context, as the experimentally observed drop of breath methane concentrations during moderate exercise cannot be explained by an altered pulmonary excretion alone. Apart from an increased dilution of breath methane within the lungs (due to a rise in the ventilation-perfusion ratio $r$), exercise will also alter the fractional (but not the absolute) perfusion of the intestine, which represents the major production site of methane in the body. This in turn leads to an additional reduction of mixed venous methane concentrations. On the basis of this rationale, a three compartment model extending the original Farhi formalism was developed and demonstrated to be in excellent agreement with measurement data obtained from a previous study as well as from a SRI-PTR-TOF-MS setup presented in this paper.

From the model equations it can be deduced that under constant resting or workload conditions the breath methane concentration $\calv(0)$ is affected by  changes of the ventilation-perfusion ratio $r$ but also by changes of the fractional intestinal blood flow $\qper$, viz.,
\begin{equation}\label{eq:var}
\calv(0)  
\approx   \frac{\qper}{r}\,\lper\, \cper(0). 
\end{equation}
This equation provides a mechanistic physiological rationale for explaining a part of the substantial intra-subject variability commonly observed in methane breath tests~\cite{wilkens1994,levitt1998}. In particular, alveolar ventilation can change considerably during breath sampling, since patients tend to hyperventilate in such a situation~\cite{cope2004}. In this context, it has been suggested to normalize breath methane concentrations with respect to CO$_2$ levels in order to improve the repeatability of breath measurements from the same individual~\cite{levitt1998}. Alternatively, as follows from 
\begin{eqnarray}
\calv(0)\qalv=  \mu\, \prm, 
\end{eqnarray} 
the present model points towards $\qalv^{-1}$, i.e., the inverse of alveolar ventilation, as an appropriate normalization factor for steady-state breath methane concentrations, as this allows for a direct estimation of the underlying endogenous methane production rate $\prm$ in the intestine. Here, $\mu \approx 0.2$ is a constant factor reflecting the expected methane loss due to flatus. 
For perspective, taking the average resting values from Figure~\ref{fig:methaneKAUN}, $\qalva=10.69$~[$\ell$/min] and $\calva(0)= (31.08-3.37)/27$~[$\mu$mol/$\ell$] yields
an estimated (resting) production rate of $54.9$~[$\mu$mol/min]. Analogously, taking average values for a workload of 75~W, $\qalva=33.12$~[$\ell$/min] and $\calva(0)= (11.92-3.37)/27$~[$\mu$mol/$\ell$] yields an estimated (workload) production rate of $52.4$~[$\mu$mol/min]. Both estimates are in good agreement with the value obtained from fitting the model dynamics to the data, see Table~\ref{table:fit}. In particular, note that the estimated endogenous methane production rate during rest and exercise is roughly constant (which is in accordance with physiological intuition), while the average breath methane concentrations during these two phases differ by a factor of roughly~2.6. This proves the efficiency of the above-mentioned normalization scheme with respect to reducing the inherent physiological variability due to Equation~(\ref{eq:var}).   

In this sense, the model is expected to contribute towards an improved comparability between breath methane measurements as well as towards a better quantitative understanding of the correlation between exhaled methane and gut methane production in general.
Measuring breath methane in combination with the present three compartment model can serve as a useful tool to assess endogenous methane production, the latter being associated with several gastrointestinal dysfunctions.

\ack
P.M., M.J., and K.U.\  gratefully acknowledge support 
from the Austrian Science Fund (FWF) under Grant No.\ P24736-B23.
A.S.\ was supported by a scholarship of {\em Aktion \"Osterreich--Ungarn}.
VR appreciates funding from the Austrian Agency for International Cooperation in Education and Research (OeAD-GmbH, project SPA 05/202 - FEM-BREATH)."
We thank the government of Vorarlberg (Austria) for its generous support.

\appendix\label{app:A}
\section{List of symbols}  
\begin{table}[H]
\centering 
\begin{tabular}{|lcc|}\hline
 {\large\strut} Parameter & Symbol & Unit\\ \hline \hline
{\large\strut}	cardiac output & $\qc$  & [$\ell$/min] \\
{\large\strut} alveolar ventilation & $\qalv$ & [$\ell$/min]\\
{\large\strut}  ventilation-perfusion ratio &  $r$ & [1]  \\
{\large\strut}	averaged mixed venous concentration & $\cven$ & [$\mu$mol/$\ell$] \\
{\large\strut}	 arterial concentration & $\cart$ & [$\mu$mol/$\ell$] \\
{\large\strut} inhaled air concentration & $\cinh$ & [ppm]\\
{\large\strut} alveolar air  concentration &  $\calv$ & [ppm]\\
{\large\strut} richly perfused compartment concentration &   $\crpt $  & [$\mu$mol/$\ell$]   \\
{\large\strut} gut compartment concentration &   $ \cper$    & [$\mu$mol/$\ell$] \\
{\large\strut}  measured exhaled concentration &  $C_{\mathrm{meas}}$  & [ppm]\\
{\large\strut} lung volume &$\valv$ & [$\ell$] \\
{\large\strut}  effective  volume of  the richly perfused compartment\ &  $\vrpt $ & [$\ell$]\\
{\large\strut}  effective  volume of  the gut compartment\ &  $\vper $ & [$\ell$]\\
{\large\strut}  metabolic rate in the richly perfused compartment& $\ml$  & [$\ell$/min] \\
{\large\strut} production rate  in the richly perfused compartment & $ \prl$  & [$\mu$mol/min]   \\
{\large\strut} production rate  in the gut compartment & $\prm  $ & [$\mu$mol/min] \\
{\large\strut} blood:air partition coefficient &  $\hen$  & [1] \\
{\large\strut}  blood:richly perfused compartment partition coefficient& $ \lrpt$    & [1]  \\
{\large\strut} blood:gut compartment partition coefficient & $ \lper$   & [1]  \\
{\large\strut} fractional blood flow to the intestine &   $\qper$   & [1] \\
{\large\strut} fractional loss of methane due to flatus &   $\mu$   & [1] \\
\hline
\end{tabular}
\end{table}

{\it Conversion from [ppb] to [nmol/$\ell$]:}\\

\noindent
A concentration of  $x$~[ppb] corresponds to  $\frac{x}{V_{m}}$ [nmol/$\ell$] 
(alternatively, $x$~[ppm] correspond to  $\frac{x}{V_{m}}$ [$\mu$mol/$\ell$]).
The molar volume $V_{m}$ can be derived from the ideal gas equation (which can safely be used for trace gases). For a measured pressure $p$ of $94600$~[Pa] and  a breath temperature  of $34$~[$^\circ$C] we  get
\begin{eqnarray}
V_{m}= \frac{ R \ T}{p} = \frac{ 8.314472 \ (273.15 +34)}{94600}=27 \ [\ell]  . \nonumber
\end{eqnarray}

\section*{References}
\bibliographystyle{amsplain}
\bibliography{methane}

\end{document}